\documentclass[saare,prev, superscriptaddress, preprintnumbers,floatset,nofootinbib,twocolumn]{revtex4-1}
\pdfoutput=1 

\pdfoutput=1
\usepackage[english]{babel}
\usepackage{bm}
\usepackage{times}
\usepackage{ulem}
\usepackage{hyperref}
\hypersetup{
    colorlinks=true,
    linkcolor=blue,
    filecolor=magenta,      
    urlcolor=blue,
    citecolor=blue,
    pdftitle={Sharelatex Example},
    pdfpagemode=FullScreen
    }
\usepackage{listings}
\usepackage{slashed}
\usepackage{color}
\usepackage{appendix}
\usepackage{mathtools}
\usepackage{epsfig}
\usepackage{dcolumn}
\usepackage{textcomp}
\usepackage{multirow}
\usepackage{graphicx}
\usepackage{soul}
\usepackage{subfigure}
\hypersetup{colorlinks=true,linkcolor={blue}}

\newcommand{\be}{\begin{equation}}
\newcommand{\ee}{\end{equation}}
\newcommand{\ba}{\begin{eqnarray}}
\newcommand{\ea}{\end{eqnarray}}

\begin{document}

\title{Constraints on Hidden Sectors Using Rare Kaon Decays}

\author{D. Cogollo}
\affiliation{Departamento de F\'{\i}sica, Universidade Federal de Campina Grande, Campina Grande, PB, Brazil}
\affiliation{Northwestern University, Department of Physics and Astronomy, 2145 Sheridan Road, Evanston, IL 60208, USA}
\author{M. J. Neves}
\affiliation{Departamento de F\'isica, Universidade Federal Rural do Rio de Janeiro,
BR 465-07, 23890-971, Serop\'edica, RJ, Brazil}
\author{T\'essio B. de Melo}
\affiliation{Millennium Institute for Subatomic Physics at High-Energy Frontier (SAPHIR), Fernandez Concha 700, Santiago, Chile}
\affiliation{Departamento de Ciencias F\'isicas, Universidad Andr\'es Bello, Sazi\'e 2212, Piso 7, Santiago, Chile}
\author{Alvaro S. de Jesus}
\affiliation{Departamento de F\'isica, Universidade Federal do Rio Grande do Norte, 59078-970, Natal, RN, Brasil}
\affiliation{International Institute of Physics, Universidade Federal do Rio Grande do Norte,  59078-970, Natal, RN, Brasil}
\author{Y. M. Oviedo-Torres}
\affiliation{Departamento de F\'isica, Universidade Federal da Paraiba, Caixa Postal 5008, 58051-970 Jo\~ao Pessoa, PB, Brasil}
\affiliation{International Institute of Physics, Universidade Federal do Rio Grande do Norte,  59078-970, Natal, RN, Brasil}
\author{F. S. Queiroz}
\email{farinaldo.queiroz@ufrn.br}
\affiliation{Millennium Institute for Subatomic Physics at High-Energy Frontier (SAPHIR), Fernandez Concha 700, Santiago, Chile}
\affiliation{Departamento de F\'isica, Universidade Federal da Paraiba, Caixa Postal 5008, 58051-970 Jo\~ao Pessoa, PB, Brasil}
\affiliation{International Institute of Physics, Universidade Federal do Rio Grande do Norte,  59078-970, Natal, RN, Brasil}

\begin{abstract}
The charged Kaon meson ($K^+$) features several hadronic decay modes, but the most relevant contribution to its decay width stems from the leptonic decay $K^+ \rightarrow \mu^+ \nu_\mu $. Given the precision acquired on the rare decay mode $K^+ \rightarrow \mu^+ \nu_\mu + X$, one can use the data to set constraints on sub-GeV hidden sectors featuring light species that could contribute to it. Light gauge bosons that couple to muons could give rise to sizeable contributions. In this work, we will use data from the  $K^+ \rightarrow \mu^+\nu_{\mu}  l^+l^-$, and $K^+  \rightarrow \mu^+ \nu_{\mu} \nu \bar{\nu}$ decays to place limits on light vector bosons present in Two Higgs Doublet Models (2HDM) augmented by an Abelian gauge symmetry, 2HDM-$U(1)_X$. We put our findings into perpective with collider bounds, atomic parity violation, neutrino-electron scattering, and polarized electron scattering probes to show that rare Kaon decays provide competitive bounds in the sub-GeV mass range for different values of $\tan\beta$. 

\noindent

\end{abstract}

\keywords{}

\maketitle
\flushbottom

\section{Introduction}
\label{sec:1}
The discovery of the Higgs boson announced by ATLAS and CMS collaborations \cite{Aad:2012tfa,CMS:2012qbp} in 2012, and a multitude of flavor physics and electroweak tests attesting the Standard Model (SM) predictions \cite{VanOnsem:2023iis,Simone:2022zir,ATLAS:2018mrn,ATLAS:2021bdj,ATLAS:2022xnu} have placed severe constraints in extended scalar sectors. The $\rho$ parameter plays an important role in this regard. Two Higgs Doublet Models (2HDM) \cite{Lee:1973iz,Haber:1984rc,Turok:1990zg,Funakubo:1993jg,Davies:1994id,Cline:1995dg} do not alter the $\rho$ parameter because the scalar doublets have hypercharge $\pm 1$ \cite{Wang:2014wua}, but they are still amenable to flavor physics data. There are several versions of 2HDM and they change according to the Yukawa lagrangian, where one decides whether and how the second Higgs doublet contributes to fermion masses \cite{Branco:2011iw}. Several phenomenological studies related to the vacuum stability \cite{Freund:1992yd,Velhinho:1994np,Nie:1998yn,Ferreira:2004yd,Battye:2011jj,Xu:2017vpq,Branchina:2018qlf,Song:2023tdg}, collider physics \cite{Aoki:2009ha,Bai:2012ex,Alves:2016bib,Barger:2013ofa,Dumont:2014wha}, and flavor physics \cite{Lindner:2016bgg,Misiak:2017bgg} have been done to explore the viable parameter space of such models. In particular, 2HDM are plagued with flavor-changing neutral interactions that severely restrict the viable parameter space of the model.
To remedy this issue, discrete symmetries have been invoked in the scalar sector, preventing the appearance of non-diagonal coupling with the scalars \cite{Branco:2011iw}. Instead of invoking arbitrary discrete symmetries to justify the contribution of only one scalar doublet to fermions masses, Abelian gauge symmetries stand as more elegant solutions. Abelian symmetries have been studied in the context of 2HDM \cite{Ko:2012hd,Ko:2013zsa,Ko:2014uka,Ko:2014uka,Crivellin:2015mga,Huang:2015wts, Wang:2016vfj,DelleRose:2017xil,Campos:2017dgc,Camargo:2018uzw}. In \cite{Ko:2012hd,Ko:2013zsa,Ko:2014uka} they are conceived to precisely solve this flavor-changing neutral current problem. In \cite{Campos:2017dgc} these flavor-changing neutral current problem was resolved and neutrino masses via a type-I seesaw mechanism were incorporated. In \cite{Camargo:2018uzw}, the type II seesaw and other possibilities were investigated. Several other phenomenological studies have been conducted in these 2HDM-$U(1)_X$ models addressing the muon anomalous magnetic moment \cite{Lindner:2016bgg,Arcadi:2021yyr}, electron-neutrino scattering \cite{Arcadi:2019uif}, and dark matter \cite{Mambrini:2015sia,Camargo:2019ukv}.   

That said, in this work, we concentrate on models that can free 2HDM from flavor-changing interactions and generate neutrino masses at the same time via the introduction of a new $U(1)_X$ gauge symmetry which is non-anomalous due to the presence of three right-handed neutrinos. The absence of flavor-changing interactions is addressed by properly assigning different quantum numbers to the two Higgs doublets. In this way, only one of the Higgs doublet contributes to fermion masses. The right-neutrinos acquire a Majorana mass term after the spontaneous symmetry breaking of the $U(1)_X$ gauge symmetry that leads to a type I seesaw mechanism due to a Dirac mass term involving the active neutrinos \cite{Minkowski:1977sc,Lazarides:1980nt,Mohapatra:1979ia,Mohapatra:1980yp,Schechter:1980gr}. The breaking of the $U(1)_X$ gauge symmetry gives rise to a hidden sector comprised of dark higgs and a vector boson. The mass of the vector boson will be proportional to the vacuum expectation value of the $U(1)_X$ breaking and the gauge couplings of the $U(1)_X$ gauge symmetry.

In principle, such 2HDM-$U(1)_X$ models can host sub-GeV hidden sectors that are entitled to an interesting phenomenology. Having in mind the historical importance of the Kaon meson to the construction of the Standard Model since its discovery in 1947 and the precision acquired in the measurement of its rare decays, several works have been put forth assessing their potential to probe hidden sectors \cite{Batell:2009jf,Davoudiasl:2012ag,Buras:2015jaq,Chiang:2016cyf,Crivellin:2016vjc}. In our work, we will focus on the sub-GeV vector boson contribution to the rare decays $K^+ \rightarrow \mu^+ \nu_\mu e^- e^+$ and $K^+ \mu^+\nu_\mu \bar{\nu} \nu$ in the context of 2HDM-$U(1)_X$. As we have a concrete and well-motivated model at hand, several phenomenological studies have been carried out in the past. Thus, we put our findings into perspective with collider bounds, atomic parity violation, neutrino-electron scattering, and polarized electron scattering probes to show that the $K^+$ meson offers an orthogonal and complementary probe to such hidden sectors.

The paper is organized as follows. In the Section \ref{sec:2}, 
we present the structure of the 2HDM models with the extra group $U(1)_{X}$ such as the particle content. The section \ref{sec:3} is dedicated to gauge bosons masses in the scenario of a light $Z^{\prime}$ and the correspondent couplings with the fermions of the particle content, relegating the detailed coupling expressions of the results to the Appendix \ref{appendix}. In Section \ref{sec:4}, we present the existing constraints on the model based on accelerators, polarized electron scattering, and neutrino-electron scattering. The section \ref{sec:5}, we derive bounds on hidden vectors using rare $K^{+}$ decays. In the section \ref{sec:6}, we discuss the final remarks and conclusions.
%
\section{The $2HDM-U(1)_X$ Model}
\label{sec:2}

Two Higgs Doublet Models (2HDM) augmented by an
abelian gauge symmetry, 2HDM-$U(1)_X$, are constructed as symmetric lagrangians under transformations of the gauge group ${\cal G}_{2HDMX} \equiv SU(3)_c\times SU(2)_L \times U(1)_{Y}\times U(1)_{X}$. As aforementioned, scalar doublets with hypercharge $Y = \pm 1$, or scalar singlets with $Y = 0$, do not modify the $\rho$ parameter, this allows us to work with a family of models featuring two scalar doublets, $\Phi_{1}$ and $\Phi_{2}$, along with a scalar singlet $\Phi_{s}$, whose scalar potential is given by:
\small
\begin{eqnarray}\label{VHiggs}
&& V(\Phi_1,\Phi_{2},\Phi_s) = m_1^{2} \left(\Phi_1^{\dag} \, \Phi_1\right)
+ m_2^{2} \left(\Phi_2^{\dag} \, \Phi_2\right) +m_{s}^{2} \left(\Phi_s^{\dag} \, \Phi_s\right) \nonumber \\
&&+\frac{\lambda_{1}}{2} \, \left(\Phi_1^{\dag} \, \Phi_1\right)^2 +\frac{\lambda_{2}}{2} \, \left(\Phi_2^{\dag} \, \Phi_2\right)^2 +\frac{\lambda_{s}}{2} \, \left(\Phi_s^{\dag} \, \Phi_s\right)^2 \nonumber \\
&&+\lambda_{3} \, \left(\Phi_1^{\dag} \, \Phi_1\right)\left(\Phi_2^{\dag} \, \Phi_2\right)
+ \lambda_{4} \left(\Phi_1^{\dag} \, \Phi_2\right)\left(\Phi_2^{\dag} \, \Phi_1\right)\nonumber \\
&& + \mu_{1} \left(\Phi_1^{\dag} \, \Phi_1\right)\left(\Phi_s^{\dag} \, \Phi_s\right)
+ \mu_{2} \left(\Phi_2^{\dag} \, \Phi_2\right)\left(\Phi_s^{\dag} \, \Phi_s\right)\nonumber \\
&& + \mu \, \Phi_{1}^{\dagger}\Phi_{2} \, \Phi_s+\mbox{h. c.} \; ,
\end{eqnarray}
\normalsize 

where the parameters $m_{i}$ $(i=1,2)$, $\lambda_{i}$ ($i=1,2,3,4$), $m_{s}$, $\lambda_{s}$, $\mu_{i}$ ($i=1,2$) and $\mu$, will be considered to be real. To avoid flavor-changing neutral interactions, all charged fermions stem from a Yukawa lagrangian where just the $\Phi_2$ doublet couples to the fermion fields,

\small
\begin{equation}
\label{2hdm_tipoI}
- \mathcal{L} _{Y _{\text{2HDM}}} = y^{d}_{i,j}  \bar{Q} _{iL} \Phi _2 d_{jR} + y^{u}_{ij}  \bar{Q} _{iL} \widetilde \Phi _2 u_{jR} + y^{e}_{ij}  \bar{L} _{iL} \Phi _2 e_{jR} + h.c., 
\end{equation}
\normalsize

being $i,j =1,2,3$ a family index,  $L_{iL}=\left(\, \nu_{iL} \;\; e_{iL} \, \right)^{T}$ left-handed doublets of leptons, $e_{iR}$ right-handed singlets of charged leptons, $Q_{iL}=\left(\, u_{iL} \;\; d_{iL} \, \right)^{T}$ left-handed doublets of quarks, $u_{iR} \, , \, d_{R} \, $  right-handed  singlets of quarks, and $\tilde{\Phi}_{2}$ a scalar doublet defined as $\tilde{\Phi}_{2}= i\,\sigma_{2} \, \Phi_{2}^{\ast}$. As for neutrinos, their masses come from a type I seesaw mechanism implemented in a Yukawa lagrangian that involves the $\Phi_{2}$ doublet, the $\Phi{_s}$ singlet, and the presence of right-handed neutrinos $(N_R)$: 

\begin{equation}
\mathcal{-L} \supset  y^{D}_{ij} \bar{L} _{iL}\widetilde \Phi _2 N_{jR} + Y^{M}_{ij}\overline{(N_{iR})^{c}}\Phi_{s}N_{Rj}\,.
\label{yukawaneutrinosphis}
\end{equation}

The three copies of right-handed neutrinos were introduced to generate neutrino masses and to free the model from gauge anomalies. Setting $q_{\Phi_1}, q_{\Phi_2}$, $q_{\Phi_s}$ as the $U(1)_X$ charges of the scalar doublets and the scalar singlet, respectively, we notice that these charges obey the relations, $q_{\Phi_1} \neq q_{\Phi_2}$, and $q_{\Phi_s}= q_{\Phi_1}-q_{\Phi_2}$ with  $u \neq -2d$ (see text in the table \ref{Table1} for more details about the $u$ and $d$ charges).

\begin{table*}[t]
\centering
\small
\begin{tabular}{|c|c|c|c|c|c|c|c|c|c|}
\hline
  &  ~$U(1)_X$~ & $U(1)_{A}$ & $U(1)_{B}$ & $U(1)_{C}$ & $U(1)_{D}$ & $U(1)_{E}$ & $U(1)_{F}$ & $U(1)_{G}$ & $U(1)_{B-L}$ \\
\hline
\hline
$L_{iL}$ & $-\frac{3}{2}(u+d)$ & $0$ & $0$ & $+3/4$ & $-3/2$ & $-3/2$ & $-3$ & $-1/2$ & $-1$ \\
$e_{iR}$ & $-2u-d$ & $-1$ & $+1$ & $0$ & $-2$ & $-1$ & $-10/3$ & $0$ & $-1$ \\
$N_{iR}$ & $-u-2d$ & $+1$ & $-1$ & $+3/2$ & $-1$ & $-2$ & $-8/3$ & $-1$ & $-1$ \\
\hline
\hline
$Q_{iL}$ &  $\frac{u+d}{2}$ & $0$ & $0$ & $-1/4$ & $+1/2$ & $+1/2$ & $+1$ & $+1/6$ & $+1/3$  \\
$u_{iR}$ &  $u$ & $+1$ & $-1$ & $+1/2$ & $+1$  & $ 0 $ & $+4/3$ & $-1/3$ & $+1/3$ \\
$d_{iR}$ &  $d$ & $-1$ & $+1$ & $-1$ & $0$ & $+1$ & $+2/3$ & $+2/3$ & $+1/3$ \\
\hline
$\Phi_1$ &  $\frac{5u}{2}+\frac{7d}{2}$ & $-1$ & $+1$ & $-9/4$ & $+5/2$ & $+7/2$ & $+17/3$ & $+3/2$ & $+2$  \\
$\Phi_{2}$ & $\frac{u-d}{2}$ & $+1$ & $-1$ & $+3/4$ & $+1/2$ & $-1/2$ & $+1/3$ & $-1/2$ & $0$  \\
$\Phi_{s}$ & $2u+4d$ & $-2$ & $+2$ & $-3$ & $+2$ & $+3$ & $+16/3$ & $+2$ & $+2$  \\
\hline
\end{tabular}
\caption{
$U(1)_X$ charges of the particles in the models. We assign a charge $q=u$ and $q=d$ for the down quarks and derive all remaining $U(1)_X$ charges by requiring that the model is free from gauge anomalies and that Eqs.\eqref{2hdm_tipoI}-\eqref{yukawaneutrinosphis} are satisfied. }
\label{Table1}
\end{table*}

After the spontaneous symmetry breaking process, the scalar fields can be parameterized as usual:
\begin{equation}
\Phi _i = \begin{pmatrix} \phi ^+ _i \\ \left( v_i + \rho _i + i\eta _i \right)/ \sqrt{2}\end{pmatrix},
\label{eqdoublet}
\end{equation}

\begin{equation}
\Phi _s = \frac{1}{\sqrt{2}} \left( v_s + \rho _s + i \eta _s \right),
\label{eqsinglet}
\end{equation}
and the charged fermions will gain Dirac masses from \eqref{2hdm_tipoI}, whereas the neutrinos will gain masses
through the type I seesaw mechanism from \eqref{yukawaneutrinosphis},

\begin{equation}
{\cal L}_{mass}^{\nu N}=\left(\nu \; N\right)
\left(\begin{array}{cc}
0 & m_D\\
m_D^T & M_R\\
\end{array}\right)\left(\begin{array}{c}
\nu \\
N \\
\end{array}\right).
\end{equation}

In the case for when $M_R \gg m_D$, being $m_D= \frac{y^D v_2}{2\sqrt{2}}$ and $M_R= \frac{y^M v_s}{2\sqrt{2}}$, it is obtained the usual light neutrino mass matrix $m_\nu = -m_D^T \frac{1}{M_R}m_D$, and the usual heavy neutrino mass matrix $m_N = M_R$. We will assume throughout that the Yukawa couplings $y^M$ are larger enough to generate right-handed neutrino masses greater than $M_{Z^\prime}$, avoiding
$Z^{\prime}$ decay into these RHN. This assumption guarantees that the $Z^\prime$ decays are into SM fermions.


In order to compute the hidden boson contribution to the $K^+$ decay, we will obtain in the next section the couplings among the $Z^{\prime}$ boson and fermions for these families of models featuring two scalar doublets, $\Phi_{1}$ and $\Phi_{2}$, along with a scalar singlet $\Phi_{s}$.

%

\section{The gauge bosons masses and couplings}
\label{sec:3}
The gauge boson masses rise from the kinetic terms of the scalar fields. To derive them, we need to correctly write the covariant derivative. The presence of a new $U(1)_X$ symmetry implies that the covariant derivative is, 

\begin{eqnarray}\label{Dmu}
D_{\mu} = \partial_{\mu}+i \, g \, T^{a} \, W_{a\mu}
+i \, g^{\prime} \, \frac{Q_Y}{2} \, \hat{B}_{\mu} + i \, g_{X} \, \frac{q_{X}}{2} \, \hat{X}_{\mu} \; ,
\end{eqnarray} 
where $g$, $g^{\prime}$ and $g_{X}$ are the dimensionless coupling constants of the $SU(2)_L , U(1)_Y , U(1)_X$ gauge groups respectively, $W_{a\mu}$ and $T^{a}=\sigma^{a}/2 \, (a=1,2,3)$ are the gauge bosons and generators of the $SU(2)_L$ group, $\hat{B}_{\mu}$ and $\hat{X}_{\mu}$ are gauge bosons of the $U(1)_{Y}$ and $U(1)_{X}$ groups respectively, $Q_Y$ (hypercharge), and $q_{X}$ are the charges associated to the groups $U(1)_{Y}$ and $U(1)_{X}$, respectively. The kinetic terms of the gauge bosons are \cite{Babu:1997st,Langacker:2008yv,Gopalakrishna:2008dv},
\small
\begin{equation}
\mathcal{L} _{\rm gauge} =  - \frac{1}{4} \hat{B} _{\mu \nu} \hat{B} ^{\mu \nu} + \frac{\epsilon}{2\, cos \theta_W} \hat{X} _{\mu \nu} \hat{B} ^{\mu \nu} - \frac{1}{4} \hat{X} _{\mu \nu} \hat{X} ^{\mu \nu}.
\label{Lgaugemix1}
\end{equation}
\normalsize
The kinetic mixing between the two Abelian groups should fulfill $\epsilon \ll 1$ to be consistent with electroweak constraints. It can be removed through the redefinition $\hat{B}_{\mu}=\eta_{X} \, X_{\mu} + B_{\mu}$
and $\hat{X}_{\mu}=X_{\mu}$ with,
\begin{eqnarray}
\eta_{X}=\frac{\epsilon/\cos\theta_{W}}{\sqrt{1-(\epsilon/\cos\theta_{W})^2}}\simeq \frac{\epsilon}{\cos\theta_{W}} \; .
\end{eqnarray}
After that, the derivative covariant operator  reads, 
\small
\begin{eqnarray}\label{DmuMod}
D_{\mu} = \partial_{\mu}+i \, g \, \frac{\sigma_{a}}{2} \, W_{a\mu}
+i \, g^{\prime} \, \frac{Q_Y}{2} \, B_{\mu} + \frac{i}{2} \, G_{X_{i}} \,
X_{\mu} \; ,
\end{eqnarray}
\normalsize
where $G_{X_i}=g_{X} \, q_{X_i}+\epsilon \, g^{\prime} \, Q_Y/\cos\theta_{W}$, with $X_i$ being the field in question. Thus, for the scalar doublets we get $G_{X_1}= g_X\, q_{\Phi_1} + \epsilon \, g^{\prime} \, Q_Y/\cos\theta_{W}$, $G_{X_2}= g_X\, q_{\Phi_2}+ \epsilon \, g^{\prime} \, Q_Y/\cos\theta_{W}$. Applying the covariant derivative Eq.\eqref{DmuMod} into the kinetic lagrangian:

\begin{eqnarray}\label{LHiggs}
{\cal L}_{Kinetic}=|D_{\mu}\Phi_1|^2+|D_{\mu}\Phi_{2}|^2+|D_{\mu}\Phi_{s}|^2 \; ,
\end{eqnarray}

and using the scalar fields in Eq.\eqref{eqdoublet}-\eqref{eqsinglet} we get, 

\small 
\begin{eqnarray}\label{Lmassgauge}
{\cal L}_{mass}=
\frac{v^2}{8}\left( \, g^2 \, W_{3\mu}W_{3}^{\;\,\mu}-2\,g\,g^{\prime}\,W_{3\mu}B^{\mu}+g^{\prime\,2}\,B_{\mu}^2\,\right)
\nonumber \\
+\frac{1}{2}\left[ \, m_{X}^2\, X_{\mu}^2
-2 \, \frac{g}{g_Z} \, \Delta^2 \, W_{3\mu}\,X^{\mu}
+2 \, \frac{g^{\prime}}{g_Z} \, \Delta^2 \, B_{\mu}\,X^{\mu} \, \right] \; ,
\end{eqnarray}
 \normalsize
where $m_{X}^2 =(G_{X_1}^2\,v_1^2+G_{X_2}^2\,v_2^2+g_{X}^2\,q_{\Phi_s}^2\,v_{s}^2)/4$, and we have defined $\Delta^2=g_Z \, (G_{X_{1}}v_1^2+G_{X_{2}}v_2^2)/4$ and $g_{Z}=\sqrt{g^2+g^{\prime2}}$. A first diagonalization process is carried out through the well-known electroweak rotation,

\begin{eqnarray}
B_\mu & =  \cos \theta _W A_\mu - \sin \theta _W Z_\mu^{0}  \nonumber\\
W_\mu ^3 & =  \sin \theta _W A_\mu + \cos \theta _W Z_\mu^{0} ,
\label{ewrotation}
\end{eqnarray}

where $\theta_{W}$ is the Weinberg angle that satisfies the relations $e=g\sin\theta_{W}=g^{\prime}\,\cos\theta_{W}$.  After the rotation \eqref{ewrotation}, the field $A_{\mu}$ is identified as the massless photon, as it must be, and the lagrangian \eqref{Lmassgauge} is reduced to, 
\small
\begin{eqnarray}\label{LmassgaugeZ0X}
{\cal L}_{mass}=\frac{1}{2} m_{Z^{0}} ^2 Z_\mu^{0}  Z^{0 \mu} - \Delta ^2 Z_\mu^{0}  X ^\mu + \frac{1}{2} m_X ^2 X_\mu X ^\mu ,
\end{eqnarray}
\normalsize
where $m_{Z^{0}}^2=g_{Z}^2\,v^2/4$, with $g_Z=g/\cos\theta_W$ and $v^{2}=v^{2}_{1}+v^{2}_{2}$. The $Z^{0}$ boson with its usual mass was nicely recovered, but there is still mixing between $Z_{\mu}^{0}$ and $X_{\mu}$. We call the attention that the mixing between these two bosons depends on $\Delta^2$ and consequently on the $U(1)_{X}$ charges of the scalar doublets. 
The diagonalization of Eq.\eqref{LmassgaugeZ0X} is carried out through the rotation,

\begin{equation}
\label{rotacao_zz_fisicos}
\begin{pmatrix} Z_\mu \\ Z ' _\mu \end{pmatrix} = \begin{pmatrix} \cos \xi & - \sin \xi \\ \sin \xi & \cos \xi \end{pmatrix} \begin{pmatrix} Z_\mu^{0} \\ X_\mu \end{pmatrix},
\end{equation}

with,

\begin{eqnarray}
\tan(2\xi) = \frac{2\Delta^2}{m_{Z^{0}}^2-m_{X}^{2}}\, ,
\label{xiangle}
\end{eqnarray}

and a new pair of eigenvalues,
\small
\begin{eqnarray}
M_{Z}^2 &=& \frac{1}{2} \left[ \, m_{Z^{0}}^2+m_{X}^2+\sqrt{ (m_{Z^{0}}^2-m_{X}^2)^2+4(\Delta^2)^2 } \, \right],
\nonumber \\
M_{Z'}^2 &=& \frac{1}{2} \left[ \, m_{Z^{0}}^2+m_{X}^2-\sqrt{ (m_{Z^{0}}^2-m_{X}^2)^2+4(\Delta^2)^2 } \right]. \nonumber\\ 
\label{eigennoapprox}
\end{eqnarray}
\normalsize

As we said, we are interested in studying the case for when the $Z^{\prime}$-gauge boson is lighter than the $Z$-boson. In this regime, the mixing angle Eq.\eqref{xiangle} can be approximated to,
\begin{equation}
\label{xi_delta}
\xi \simeq \frac{\Delta ^2}{m^2_Z}=\dfrac{1}{g_{z}}(G_{X1}\cos^{2}\beta+G_{X2}\sin^{2}\beta),
\end{equation}which simplifies to,
\begin{equation}
\xi \simeq \epsilon_Z + \epsilon \tan \theta _W ,
\label{xisimplifiedEq}
\end{equation}where,
\begin{equation}
\epsilon_Z = \dfrac{g_{X}}{g_{Z}}(q_{\Phi_1}\cos^{2}\beta+q_{\Phi_2}\sin^{2}\beta).
\end{equation}

Using this small mixing approximation the eigenvalues \eqref{eigennoapprox} reduce to,
\begin{eqnarray}
\label{massesZZ'}
M_{Z}^2 & \simeq & \frac{1}{4} \, g_{Z}^2 \, v^{2}, \;\nonumber\\
M_{Z^{\prime}} & \simeq & \frac{1}{2} \, g_{X} \, |q_{\Phi_s}| \, \sqrt{ \, v_{s}^{2}
+ v^2 \, \sin^2\beta \, \cos^2\beta \, }.
\end{eqnarray}where we have used the common parameterization of the vacuums with $v_1=v\,\cos\beta$, and $v_2=v\,\sin\beta$, such that $\tan\beta=v_2/v_1$. Realize that $M_{Z^{\prime}}$ in Eq.\eqref{massesZZ'} depends on the free parameters $g_X$, $v_{s}$, $\beta$ and the $U(1)_X$ charge of the singlet scalar.

We have derived the physical gauge bosons, their masses and the relevant mixings. These quantities are key to our findings, which are governed by the neutral current of the hidden vector boson. The full calculation of this neutral current is presented in the Appendix \ref{appendix}, as well as the neutral current mediated by the standard $Z$ boson. For now, we just show here the general expression we used to derive our results,

\begin{eqnarray}\label{LintZp}
\,{\cal L}_{Z^{\prime}}^{\, int}= g_{V}^{(\Psi_{i})} \, \, \overline{\Psi_{i}} \, \slashed{Z}^{\prime} \, \Psi_{i}
+ g_{A}^{(\Psi_{i})} \, \, \overline{\Psi_{i}} \, \slashed{Z}^{\prime} \, \gamma_{5} \, \Psi_{i} \; ,
\end{eqnarray}

being $\Psi_{i}$ each one of the fermions of the SM, $g^{V}$ and $g^{A}$ the vector and axial factors whose explicit forms are presented in the Appendix \ref{appendix}

\section{Existing constraints}
\label{sec:4}

In this section, we discuss the existing bounds for the region of interest, namely 10~MeV$ < M_{Z^{\prime}}<$200MeV. We will discuss collider limits, polarized electron scattering, and neutrino-electron scattering. We start with collider searches for charged scalars which turn out to be the most relevant one. 

\subsection{Collider Bounds on Charged Scalars}

\begin{figure}[t]
\includegraphics[scale=0.48]{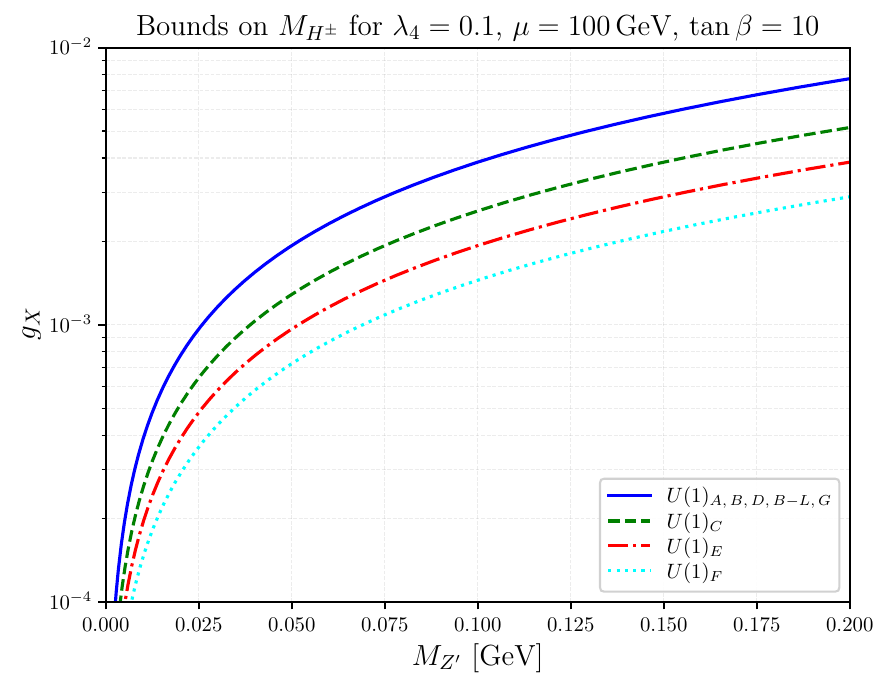}
\includegraphics[scale=0.48]{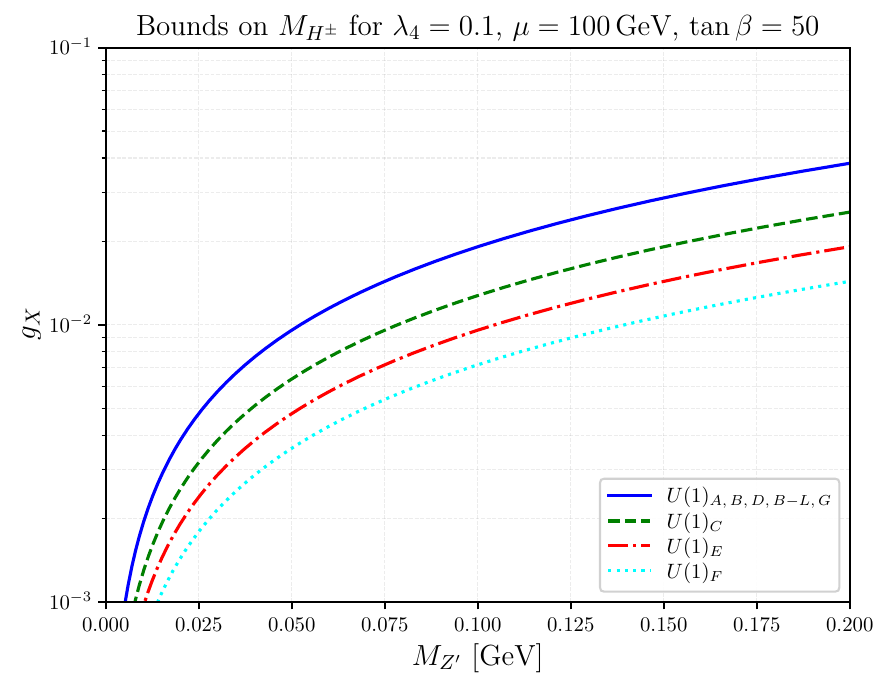}
\caption{Summary of the limits for all of the models in the $g_X \times m_{Z^\prime}$ parameter space coming from the experimental limits on the mass of the charged Higgs $H^{+}$ searches at LEP.}
\label{fig:boundsmhpLEP}
\end{figure}

An important bound rises from charged Higgs boson  searches, $H^{+}$. The most relevant bound in our work is the one derived from LEP data, which has presented a lower bound on  $M_{H^{\pm}}$ ($M_{H^{\pm}} > 80$ GeV for $\tan{\beta}\geq 10$) for a Type-I 2HDM charged Higgs boson $H^{+}$ \cite{LEP}. For the family of $SU(2)_{L} \times U(1)_{Y} \times U(1)_{X}$ 2HDM extensions of the Standard Model we are studying here, $M_{H^{\pm}}$ is given by \cite{CamargoPLB} ,
\begin{eqnarray}
M_{H^{\pm}}^{2}=\left( \sqrt{2} \, \mu \, v_s-\lambda_{4} \, v_1 \, v_2 \right)\frac{v^2}{2v_1\,v_2}
\end{eqnarray}which implies that,
\begin{eqnarray}\label{mHp}
M_{H^{\pm}}^{2} = \frac{\sqrt{2}\mu v_s - \lambda_4 v^2\sin{\beta}\cos{\beta} }{2\sin{\beta}\cos{\beta}}\; ,
\end{eqnarray}
depending on the $\beta$-parameter, the VEV $v_s$ of the singlet scalar, as well as the parameters $\mu$ and $\lambda_4$ present in the scalar potential of the model, Eq.\eqref{VHiggs}. For the parameters $\mu$ and $\lambda_4$ of the scalar potential, we must be careful in order to avoid an imaginary $M_{H^{\pm}}$ value, and pertubativity violation, respectively. Doing so, we assumed reasonable values for these parameters in our calculations, with $\mu=100$ GeV ($\mu$ is only bounded from below) and $\lambda_4=0.1$ ($\lambda_4$ parameter only being bounded from above). We performed our analysis for two different values of $\tan{\beta}$, $\tan{\beta} = 10$ and $\tan{\beta} = 50$. After setting these values of $\tan{\beta}$, $M_{H^{\pm}}$ in Eq.\eqref{mHp} depends just on $v_s$. The recast of this bound on the $g_X, M_{Z^\prime}$ plane is made by exploring the mutual dependence of the mass of the charged Higgs and $Z^\prime$ with the free parameter $v_s$, as seen in Eq.\eqref{massesZZ'} and Eq.\eqref{mHp}. Being $q_{\Phi_s}$ the scalar singlet charge under $U(1)_X$, and as $M_{Z^\prime}$ depends on $|q_{\Phi_s}|$, this bound becomes model-dependent, as it can be seen in Figure \eqref{fig:boundsmhpLEP} for the cases of $\tan{\beta}=10$ (top) and $\tan{\beta}=50$ (bottom), with the area above the lines being excluded since they represent a region of the parameter space where the mass of the charged Higgs is below the bound from LEP. There is a similarity among the bounds of the models $U(1)_{A}$, $U(1)_{B}$, $U(1)_{D}$, $U(1)_{B-L}$ and $U(1)_{G}$, which comes from the fact that these models have $|q_{\Phi_s}|=2$, while the models $U(1)_{C}$, $U(1)_{E}$, and $U(1)_{F}$ show stronger bounds due to larger values of $|q_{\Phi_s}|$.

\subsection{Polarized Electron Scattering}

Low energy polarized electron scattering offers an orthogonal probe to light hidden particles that feature kinetic and mass mixing with the Z boson.
In other words, the weak currents in the presence of such mass mixings and kinetic mixings can be parameterized in terms of a shift in the Weinberg angle as follows \cite{Arcadi:2019uif},
\begin{equation}
        \Delta\sin^{2}\theta_{W}\simeq -0.42\epsilon\delta\frac{m_{Z}}{m_{Z^\prime}}f(Q^{2}/m_{Z^\prime}^{2}) \; ,
\end{equation}
where $f(Q^{2}/m_{Z^\prime}^{2})$ is the propagator function \cite{Arcadi:2019uif} with the masses and the energy $Q$ in MeV units. $\delta$ is the mass mixing parameter defined as,
\begin{equation}
    \delta = \frac{m_Z}{m_{Z^\prime}}\epsilon_Z \; .
\end{equation}
 Thus, if we happen to measure the Weinberg angle at a given energy $Q$, we can limit the contribution of new physics to $\sin^2\theta_W$ as a function of $\delta$ and $\epsilon$ with, 
\begin{equation}
\epsilon^{2} \simeq \frac{5.7}{\delta^{2}} (\Delta\sin^{2}\theta_{W})^{2}\left(\frac{m_{Z'}^{2}+Q^{2}}{m_{Z}m_{Z'}}\right)^{2}.
\end{equation}

The E158 experiment measured $\sin^2\theta_W =0.2329(13)$ with $Q=160$~MeV \cite{PDG}, which leads to,
\begin{equation}
\epsilon^{2}<\frac{1.5 \times 10^{-5}}{\delta^{2}}\left(\frac{m_{Z'}^{2}+160^{2}}{m_{Z}m_{Z'}}\right)^{2}.
\end{equation}
For the $U(1)_{B-L}$ model $\epsilon_Z = 2 g_X \cos^2\beta/g_Z$. Consequently,
\begin{equation}
\epsilon < 1.8 \times 10^{-5}/g_X \; ,
\end{equation} 
for $m_{Z^\prime}\sim 100$~MeV, which is the region of interest here. A similar relation can be derived for the other models. The bounds arising from rare Kaon decay will lie in the range of $g_X \sim 10^{-2}- 10^{-3}$, we assume the kinetic mixing to be sufficiently small ($\epsilon \sim 10^{-4})$ to be consistent with polarized electron scattering measurements.

\begin{figure}[t!]
\includegraphics[scale=0.48]{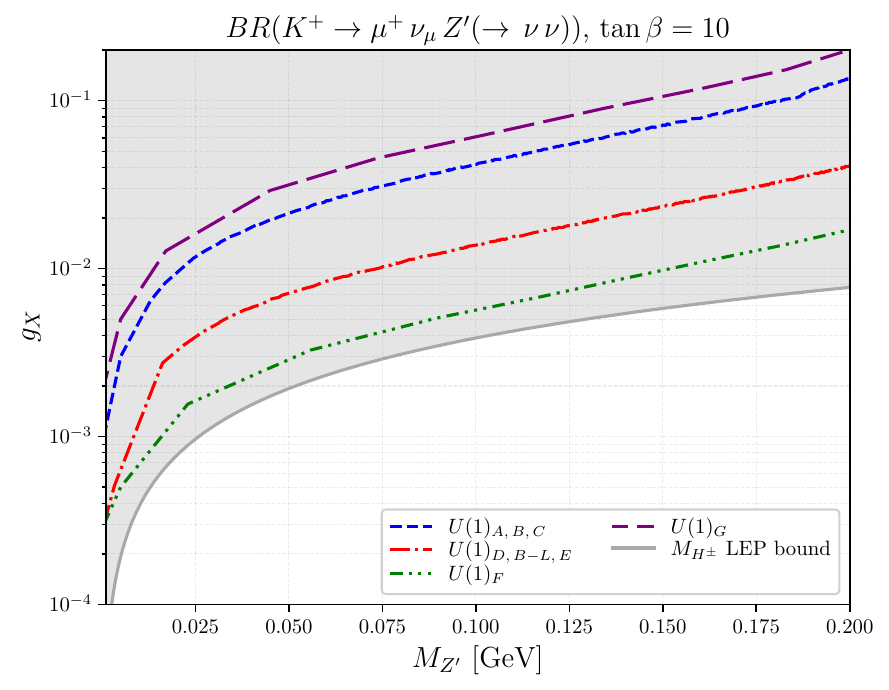}
\includegraphics[scale=0.48]
{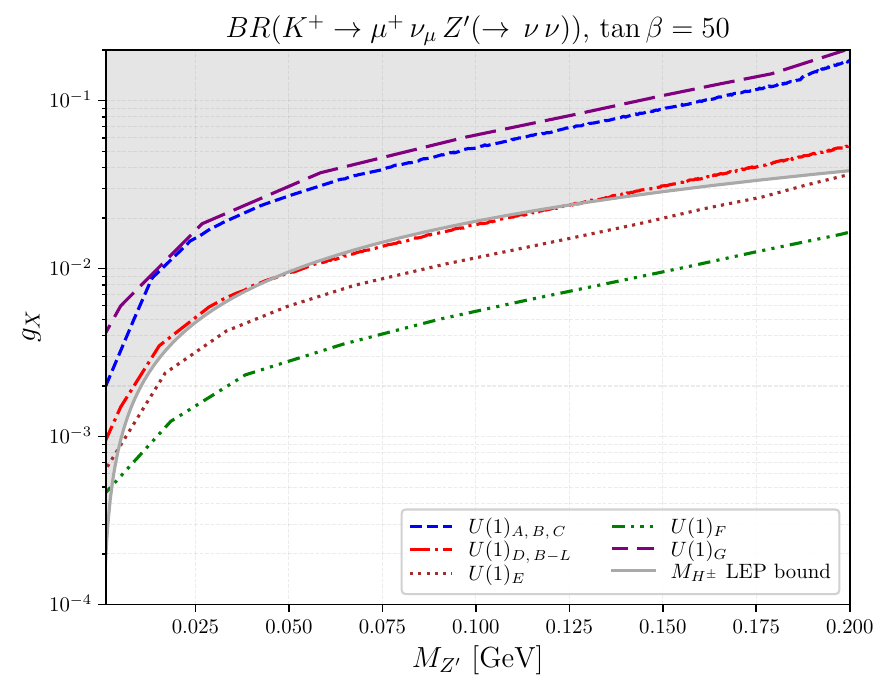}
\caption{Bounds on $g_X$ as a function of the hidden vector mass for $\tan\beta=10$ and $\tan\beta=50$, respectively, using the $K^+$ decay into $\mu^+ \nu_\mu \bar{\nu}\nu$. The LEP bound delimited by a grey region stems from charged scalar searches. As the charged scalar mass is related to $\tan\beta$, the LEP bound changes for different values of $\tan\beta$. From the plots, we conclude that $K^+$ decays yield competitive limits, especially for $\tan\beta=50$.}
\label{fig:boundsBRtan10}
\end{figure}

\subsection{Neutrino-Electron Scattering}

Hidden particles can contribute to neutrino-electron scattering. Using data from TEXONO, CHARM-II and GEMMA collaborations, one can compute the neutrino-electron scattering cross section given by,

\begin{equation}
   \frac{d\sigma}{dT} (\bar{\nu}e\rightarrow \bar{\nu}e) =\frac{m_e G_F^2}{4\pi}\left[ g_1^2+g_2^2\left(1-\frac{T}{E_\nu}\right)^2 \! -g_1g_2 \frac{m_e T}{E_\nu^2}\right] , 
   \label{eqsigmanue}
\end{equation} 
where $E_\nu$ and T are the neutrino energy and electron recoil energy, respectively. A similar cross-section exists for $\nu e \rightarrow \nu e$ scattering, and the scattering cross-section is equal to Eq.\eqref{eqsigmanue}, by interchanging $g_1$ and $g_2$. The couplings $g_{1,2}$ encode all the coupling constants involved in the SM and $Z^\prime$ interactions. For heavy hidden vectors with masses larger than 1.5GeV, the neutrino-electron scattering can be described in terms of effective operators. The strongest constraint in this case arises from CHARM-II, which features the best precision. For hidden vectors with masses between $400$~keV and $1.5$~GeV a trade-off between energy threshold and precision takes place. Considering data from GEMMA, TEXONO, and CHARM-II that cover various energy scales, constraints on the kinetic mixing parameter were derived from several vector bosons \cite{Lindner:2018kjo}. In particular, for $M_{Z^\prime} =1$~MeV, we need to impose $\epsilon < 10^{-4}$, whereas for $M_{Z^\prime} = 100$~MeV, we require $\epsilon < 10^-3$. In our work, we are focusing on $10\,{\rm MeV} < M_{Z^\prime} < 200$~MeV, thus we safely obey these limits by assuming $\epsilon=10^{-4}$. We have checked that values of $\epsilon \ll 10^{-4}$ yield no meaningful change in our results. 

Taking into account three different sources of constraints, we conclude that the neutrino-electron scattering and collider searches for charged scalars provide the most restrictive bound in our model. The latter will be shown in our figures, whereas the former is obeyed by assuming $\epsilon=10^{-4}$ throughout.

In summary, we have reviewed different constraints that are relevant to our model, and concluded that as long as we keep the kinetic mixing to be $\epsilon-10^{-4}$, the bound on the charged scalar rising from LEP data ends up being the most relevant for our reasoning.  

\section{Bounds from Rare $K^{+}$ decays}
\label{sec:5}

Rare decays of the $K^{+}$ meson can be considered as radiative corrections to its main decay mode $K^{+} \longrightarrow \mu^{+}\nu_{\mu}$, when from the muon leg is radiate a light $Z^{\prime}$ boson that subsequently decay into $Z^{\prime}\longrightarrow e^{-}e^{+}$ or $Z^{\prime}\longrightarrow \bar{\nu}\nu$. In both cases the first information we should get is the width of the decay mode $K^{+}\rightarrow \mu^{+} \, \nu_{\mu} \, Z^{\prime}$. Recently, it was shown in \cite{Datta} that its analytical expression is given by,
\small
\begin{eqnarray}\label{GammaKZp}
\Gamma(K^{+}\rightarrow \mu^{+} \, \nu_{\mu} \, Z^{\prime})=\frac{1}{64\pi^3M_{K}}
\int_{E^{min}}^{E^{max}}
\sum_{spins}|{\cal M}|^2 \, dE_{\nu} \, dE_{\mu},\nonumber\\
\end{eqnarray}
\normalsize
where $|\mathcal{M}|^2$ is the square of the invariant amplitude of the process and $M_{K}=493.677$ MeV is the mass of the meson $K^{+}$. The limits of the integration are written as,

\begin{eqnarray}
E_{\mu}^{min} &=& m_{\mu} \; ,
\nonumber \\
E_{\mu}^{max} &=& \frac{M_{K}^2+m_{\mu}^2-M_{Z^{\prime}}^2}{2\,M_{K}} \; ,  
\nonumber \\
E_{\nu}^{min} &=& \frac{ M_{K}^2+m_{\mu}^2-M_{Z'}^2-2M_{K}\,E_{\mu} }{2\,(M_{K}-E_{\mu}+\sqrt{E_{\mu}^2-m_{\mu}^2})} \; ,
\nonumber \\
E_{\nu}^{max} &=& \frac{M_{K}^2+m_{\mu}^2-M_{Z'}^2-2M_{K}\,E_{\mu}}{2\,(M_{K}-E_{\mu}-\sqrt{E_{\mu}^2-m_{\mu}^2})} \; ,
\end{eqnarray}

in which $m_{\mu}=105.66$ MeV is the mass of the muon, $E_{\mu}$ is the energy in the final state of the muon,
constrained by the condition $E_{\mu}\geq m_{\mu}$, and $E_{\nu}$ is the energy of the outgoing neutrino, bounded by kinematic conditions. The full expression of $|\mathcal{M}|^2$ in Eq.\eqref{GammaKZp} is \cite{Datta} ,
%
%
\begin{widetext}
\begin{eqnarray}
\sum_{spins}|{\cal M}|^{2}=\frac{G_{F}^2\,f_{K}^2\,|V_{us}|^2}{M_{Z^{\prime}}^2(Q^2-m_{\mu}^2)^2}
\left[ \, g_{L}^2\,Q^4 (2E_{\mu}M_{K}(M_{Z'}^2+m_{\mu}^2-Q^2)-2E_{Z'}M_{K}M_{Z'}^2
\right.
\nonumber \\
\left.
-M_{K}^2\,m_{\mu}^2+M_{K}^2\,Q^2-m_{\mu}^4+m_{\mu}^2\,Q^2+2M_{Z'}^4-m_{\mu}^2\,M_{Z'}^2-M_{Z'}^2\,Q^2)
\right.
\nonumber \\
\left.
+ 6 \, g_{L} \, g_{R} \, m_{\mu}^2 \, M_{Z'}^2\,Q^2(Q^2-M_{K}^2)+g_{R}^2\,m_{\mu}^2(-2E_{\mu}M_{K}^3(M_{Z'}^2+m_{\mu}^2-Q^2)
\right.
\nonumber \\
\left.
+2E_{Z'}(E_{\nu}M_{Z'}^2(M_{K}^2-Q^2)
+M_{K}^3(Q^2-m_{\mu}^2)+M_{K}Q^2(M_{Z'}^2+m_{\mu}^2-Q^2))
\right.
\nonumber \\
\left.
+ M_{K}^4\,m_{\mu}^2+M_{K}^4\,M_{Z'}^2-M_{K}^4\,Q^2 + M_{K}^2\,m_{\mu}^4-M_{K}^2\,m_{\mu}^2\,Q^2
\right.
\nonumber \\
\left.
-3\,M_{K}^2\,M_{Z'}^4+2\, M_{K}^2 \, m_{\mu}^2 \, M_{Z'}^2+M_{Z'}^4\,Q^2-m_{\mu}^2\,M_{Z'}^2\,Q^2 ) \, \right] \; ,
\end{eqnarray}
\end{widetext}
where $g_{L}=g_{V}-g_{A}$ and $g_{R}=g_{V}+g_{A}$ are defined as combinations of the vector $(g_V)$ and axial $(g_{A})$
coefficients from Eq.\eqref{LintZp}, $Q^2=M_{K}^2-2M_{K}E_{\nu}$, $E_{Z'}$ is the
energy of the emitted $Z^{\prime}$ boson,
$G_{F}=1.166 \times 10^{-5} \, \mbox{GeV}^{-2}$ is the Fermi constant, $f_{K}=155.7$ MeV the Kaon decay constant, and
$|V_{us}|=\sin\theta_{c}=0.22534$ is the sine of the Cabibbo angle.

\begin{figure}[t]
\includegraphics[scale=0.48]
{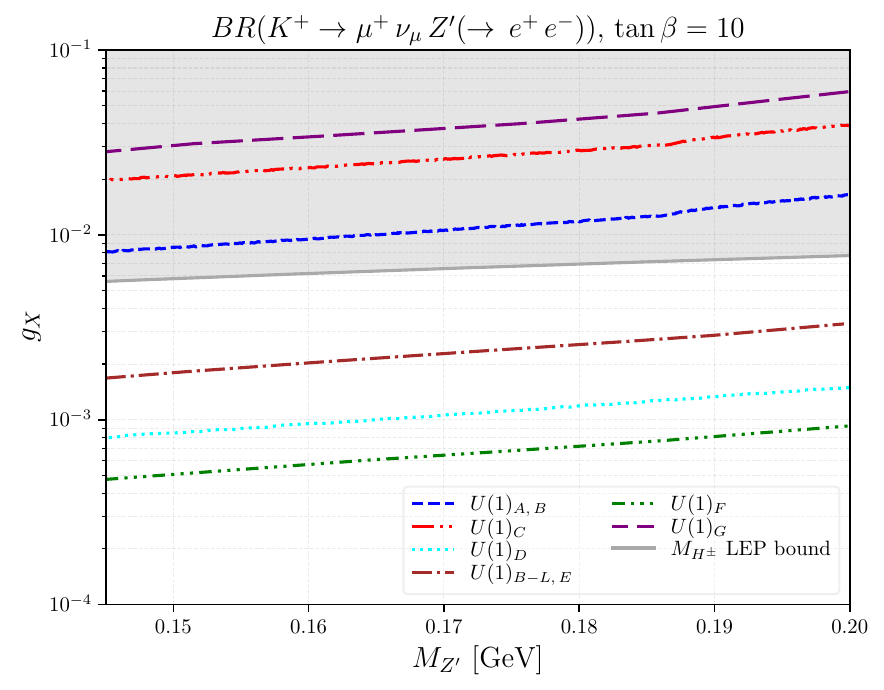}
\includegraphics[scale=0.48]{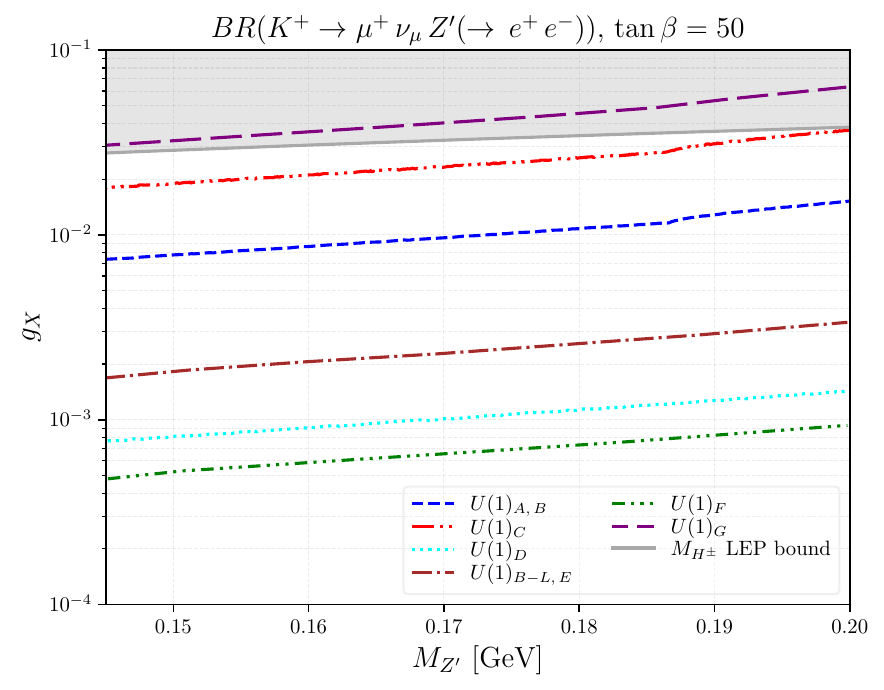}
\caption{Bounds on the $g_X$ as a function of the hidden vector mass, assuming $\tan{\beta}=50$ for different $U(1)_X$ extensions. Our upper limits are represented with dotted and dashed curves. The LEP bound is delimited by the grey region. Therefore, the impressive precision achieved in this decay mode turned it into an excellent laboratory for hidden vector bosons. From the plots, we find that it provides the most constraining bounds for several models for either $\tan\beta=10$ or $\tan\beta=50$.}
\label{fig:boundsBRtan50}
\end{figure}

\subsection{The $K^{+}\rightarrow \mu^{+} \, \nu_{\mu} \, Z^{\prime} \,(\longrightarrow\,\overline{\nu}\,\nu\,)$ decay}

NA62 collaboration reported null results for the rare decay $K^+ \rightarrow \mu^+ \nu_\mu \bar{\nu}\nu$, which led to the 90$\%$ CL upper limit \cite{NA62} ,
\begin{eqnarray}
\label{Kinvisible}
{\cal B}(K^{+}\rightarrow \mu^{+} \, \nu_{\mu} \, \overline{\nu}\,\nu) < 1.0 \times 10^{-6} \;.
\end{eqnarray}

When $Z^{\prime}$ is on-shell, we can write Eq.\eqref{Kinvisible} as,

\begin{eqnarray}
\label{Kinvisible2}
{\cal B}(K^{+}\rightarrow \mu^{+} \, \nu_{\mu} \, Z^{\prime}) \cdot {\cal B}(Z^{\prime}\rightarrow \overline{\nu}\,\nu) < 1.0 \times 10^{-6}\; ,
\end{eqnarray}

with each one of the branching in Eq.\eqref{Kinvisible2} defined as usual,

\begin{subequations}
\begin{eqnarray}
{\cal B}(K^{+}\rightarrow \mu^{+} \, \nu_{\mu} \, Z^{\prime}) &=& \frac{\Gamma(K^{+}\rightarrow \mu^{+} \, \nu_{\mu} \, Z^{\prime})}{\Gamma_{K^{+}}} \; ,
\\
{\cal B}(Z^{\prime}\rightarrow \overline{\nu}\,\nu) &=& \frac{\Gamma(Z^{\prime}\rightarrow \overline{\nu}\,\nu)}{\Gamma_{Z^{\prime}}}  \; ,
\end{eqnarray}
\end{subequations}
being $\Gamma_{K^{+}}=5.38 \times 10^{-17}$ GeV \cite{PDG} the  total decay width of the meson $K^{+}$ and $\Gamma_{Z^{\prime}}$ the total decay width of the $Z^{\prime}$ boson. The width decay of the $Z^{\prime}$ boson into any pair of anti-fermion-fermion $(\,\overline{f} \, f\,)$ is given by,

\begin{eqnarray}
\Gamma_{Z^{\prime}\longrightarrow \bar{f}f}=\frac{M_{Z^{\prime}}}{12\pi}N_{c}^{f} \;
\sqrt{1-\frac{4m_{f}^2}{M_{Z^{\prime}}^2} } \biggl[ (g_{V}^{f})^2+(g_{A}^{f})^2 \nonumber\\ +\frac{2m_{f}^2}{M_{Z^{\prime}}^2}
\left( (g_{V}^{f})^2-2\,(g_{A}^{f})^2 \right)  \biggr],
\label{GammaZpartial}
\end{eqnarray}
where $N_{c}^{f}$ is the number of colors of the final state fermion, $N_{c}^{Q} = 3$ for quarks, $N_{c}^{\ell} = 1$
for charged leptons and neutrinos. The partial width decay Eq.\eqref{GammaZpartial} is allowed for when $M_{Z^{\prime}} > 2\,m_{f}$, and the total width decay $\Gamma_{Z^{\prime}}$ is the sum over each one of the kinematically allowed partial width decay $\Gamma_{Z^{\prime}\longrightarrow \bar{f}f}$ in Eq.\eqref{GammaZpartial}.
\subsection{The $K^{+}\rightarrow \mu^{+} \, \nu_{\mu} \, Z^{\prime} \,(\longrightarrow\,e^{+}\,e^{-}\,)$ decay}
However, the search for the rare decay mode of the type $K^{+}\rightarrow \mu^{+} \, \nu_{\mu} e^{+} e^{-} $ resulted in the recent observation of over two thousand events. Hence, instead of an upper limit the branching ratio was measured to be \cite{ee},
\begin{eqnarray}
\label{Kepositron}
{\cal B}(K^{+}\rightarrow \mu^{+} \, \nu_{\mu} \, e^{-}e^{+}) = (7.06 \pm 0.31) \times 10^{-8} \; ,
\end{eqnarray}
It is important to point out that Eq.\eqref{Kepositron} was derived for an invariant electron positron mass $m_{ee} > 145$ MeV. Thus, it applies for $m_{Z^{\prime}} > 145$ MeV. Expressing the branching ratio as a product of the branchings,
\small
\begin{equation}
\label{Kepositron2}
{\cal B}(K^{+}\rightarrow \mu^{+} \, \nu_{\mu} \, Z^{\prime}) \cdot {\cal B}(Z^{\prime}\rightarrow e^{-}e^{+}) = (7.06 \pm 0.31) \times 10^{-8} \; ,
\end{equation}
\normalsize
with,
\begin{eqnarray}
{\cal B}(Z^{\prime}\rightarrow e^{-}e^{+}) &=& \frac{\Gamma(Z^{\prime}\rightarrow e^{-}e^{+})}{\Gamma_{Z^{\prime}}}  \; ,
\end{eqnarray}we can place constraints on the properties of the hidden vector.

\subsection{Bounds on Hidden Sectors}

We remind the reader that we are interested in deriving bounds on hidden sectors using data from rare $K^+$ decay. Due to kinematics, we will focus on the $M_{Z^\prime}=1-200$~MeV mass range. Regarding the possible $Z^\prime$ decays, the $Z^\prime$ may decay into the first generation of leptons and light quarks through Eq.\eqref{LintZp}. We highlight that the hadronic contributions have been calculated at parton level, and right-handed neutrinos are taken to be heavy, thus not kinematically accessible.  As the branching ratios depend on the couplings between the $Z^\prime$ with fermions which in term are governed by $\tan\beta$ and $g_X$ for a given $U(1)_X$ gauge symmetry, we can plot the constraints in the $g_X$ vs $M_{Z^\prime}$ plane. In the Appendix \ref{appendix}, we show the explicit expression of the neutral current.

Therefore, we derive our numerical results for $\tan_{\beta}=10$ and $\tan_{\beta}=50$, and overlay our findings with LEP bound on the singly charged scalar discussed previously, which happens to be the most restrictive one as we are assuming the kinetic mixing to be sufficiently suppressed. The LEP bound is delimited by a grey region. In Fig.\ref{fig:boundsBRtan10} we present our lower bounds on the gauge couplings as a function of the $Z^\prime$ mass based on the $K^{+}\rightarrow \mu^{+} \, \nu_{\mu} \, Z^{\prime} \,(\longrightarrow\,\overline{\nu}\,\nu\,)$ decay. In the upper panel, where   $\tan_{\beta}=10$, we clearly see that the search for rare $K^+$ decays results in limits that are weaker than those from LEP, which refers to $M_{H^{+}} < 80$ GeV. Analyzing the plots, we find that all the $U(1)_{X}$ models are excluded by LEP for the case of $\tan{\beta}=10$, however, increasing the $\beta$-parameter for
$\tan\beta=50$, we observe that the models $U(1)_{E}$ and $U(1)_{F}$ are not ruled out by the LEP bound, showing that the search for rare meson decays can produce stronger bounds for these models. This happens due to the fact that the LEP bound gets weaker (see \autoref{fig:boundsmhpLEP}) while the bounds on ${\cal BR}(Z^\prime \to \nu \nu)$ do not change by much with higher values of $\tan{\beta}$, with the only exception being the model $U(1)_{F}$, which presents a change by a factor of 2.

In the bottom panel, with $\tan\beta=50$, the situation changes significantly though, especially for models in which the invisible decay is large, such as $U(1)_{B-L}$,$ U(1)_E$, $U(1)_F$, etc. With $\tan\beta=50$, we impose $g_X< 1 \times 10^{-3}$ for $M_{Z^\prime} \sim 25$~MeV for the $U(1)_F$ model, and $g_X< 5 \times 10^{-3}$ for the $U(1)_{B-L}$ model.  We remind the reader that we assumed $\epsilon=10^{-4}$ throughout, but we have checked that our numerical results will not change by a factor of two if other values of $\epsilon$ are assumed. 

Regarding the $K^{+}\rightarrow \mu^{+} \, \nu_{\mu} \, Z^{\prime} \,(\longrightarrow\, e^+\,e^-)$ decay, the constraints in the parameter space of the model are derived enforcing the hidden vector contribution to be smaller than the error bar in Eq.\eqref{Kepositron2}. In this case, we realize that the study of rare $K^+$ decays becomes quite fruitful, as seen in Fig.\ref{fig:boundsBRtan50}. In both cases, with $\tan\beta=10$ and $\tan\beta=50$, the bounds we get from Kaon decays exclude a much larger region of parameter space than those from collider searchers. In particular, for the $U(1)_{B-L}$ model we find $g_{X}< 2 \times 10^{-3}$ across the entire parameter space, while for the $U(1)_F$ model, $g_X < 5 \times 10^{-4}$.  The constraints for the other models can be easily extracted from Fig.\ref{fig:boundsBRtan50} which resulted to be the most restrictive one.

In summary, we conclude that the $K^{+}\rightarrow \mu^{+} \, \nu_{\mu} \, Z^{\prime} \,(\longrightarrow\,\overline{\nu}\,\nu\,)$ rare decay does not lead to very restrictive bounds unless large values of $\tan\beta$ are adopted. Whereas, for $K^{+}\rightarrow \mu^{+} \, \nu_{\mu} \, Z^{\prime} \,(\longrightarrow\, e^+\,e^-)$, which has a branching ratio over an order of magnitude smaller than the $K^{+}\rightarrow \mu^{+} \, \nu_{\mu} \, Z^{\prime} \,(\longrightarrow\,\overline{\nu}\,\nu\,)$, the probe of hidden vector becomes rewarding. For several $U(1)_X$ models, across the entire region of interest, the $K^+$ decay into  $\mu^{+} \, \nu_{\mu} \, e^+\,e^-)$ produces stronger limits than those from collider searches.

\section{Conclusions}
\label{sec:6}

Kaon mesons have played a key role in the construction of the Standard Model since their discovery in cosmic rays in 1947. They were of paramount importance to the understanding of the charged currents with the observation of the $K^{+} \rightarrow \mu^+ \bar{\nu}$ decay and were also essential in establishing the foundations of CP violation in 1964. Several studies to precisely measure the kaon decays have been conducted since then. Recently, a thousand of excess events were observed, leading to the measurement of the $K^+$ decay into $\mu^+ \nu_\mu e^+e^-$. Motivated by this, we derived constraints on hidden vectors that belong to Abelian gauge symmetries in the context of Two Higgs Doublet Models. Our findings
are based on rare $K^+$ decays into $\mu^+ \bar{\nu}_\mu \nu \bar{\nu}$ and $\mu^+ \bar{\nu}_\mu e^+ e^-$.

Putting our results into perspective with other existing limits, we concluded that the most constraining rises from LEP searches for charged scalars. As the mass of the charged scalar is impacted by $\tan\beta$, we recast this limit as we explored different regions of the parameter space. 

In summary, we found that the $K^+$ decay into $\mu^{+} \, \nu_{\mu}\, \overline{\nu}\,\nu$ rare decay is not very constraining, except when larger values of $\tan\beta$ are assumed. Nevertheless, the measurement of the $K^{+}\rightarrow \mu^{+} \, \nu_{\mu} \, e^+\,e^-$ decay significantly improved the power of probing hidden vectors. This decay model gave rise to the strongest limit for several $U(1)_X$ symmetries, for both $\tan\beta=10$ and $\tan\beta=50$. Conclusively, rare kaon decays constitute a great laboratory for probing light hidden particles.

\acknowledgments

The authors thank Carlos Pires and Yoxara Villamizar for discussions. D.C thanks André de Gouvêa for the useful discussions, and motivation to develop this work. This work was financially supported by Simons Foundation (Award Number:1023171-RC), FAPESP Grant 2021/01089-1, ICTP-SAIFR FAPESP Grants 2021/14335-0, CNPq Grant 307130/2021-5, FONDECYT Grant 1191103 (Chile) and ANID-Programa Milenio-code ICN2019\_044.\\

\section{Appendix}
\label{appendix}
%


%
%
%

Here, we will obtain the interactions among fermions and the $Z^\prime$ field using the kinetic term,  

\begin{equation}
\mathcal{L} _{\text{fermion}} = \sum _{\text{férmions}} \bar{\Psi} ^L i \gamma ^\mu D_\mu \Psi ^L + \bar{\Psi} ^R i \gamma ^\mu D_\mu \Psi ^R ,
\label{ldiracu1}
\end{equation}

The covariant derivative must be written as a function of the physical neutral gauge bosons to then substitute it into Eq.\eqref{ldiracu1}. Doing so, and after a big algebra, we obtained for the left-handed fields,

\begin{widetext}
\begin{equation}
\begin{split}
\bar{\Psi} ^L i \gamma ^\mu D_\mu ^L \Psi ^L& = - e Q_f \bar{\psi} _f ^L \gamma ^\mu \psi _f ^L A_\mu \\
&- \left[ g_Z \left( T^L _{3f} - Q_f \sin ^2 \theta _W \right) \cos \xi - \frac{1}{2} \left( \epsilon g _Z Q_{Yf} ^L \tan \theta _W + g_X Q_{Xf} ^L \right) \sin \xi \right] \bar{\psi} _f ^L \gamma ^\mu \psi _f ^L Z_\mu \\
&- \left[ g_Z \left( T^L _{3f} - Q_f \sin ^2 \theta _W \right) \sin \xi + \frac{1}{2} \left( \epsilon g _Z Q_{Yf} ^L \tan \theta _W + g_X Q_{Xf} ^L \right)\cos \xi \right] \bar{\psi} _f ^L \gamma ^\mu \psi _f ^L Z' _\mu,
\end{split}
\label{diracLu1}
\end{equation}
\end{widetext}
where were used the relations $g \sin \theta _W = g' \cos \theta _W = e$, $g_Z = g / \cos \theta _W$, $g ' = g_Z \sin \theta _W$ and $T^3 + Q_Y / 2 = Q$. As for the right-handed fields, it is enough to replace $T^L _{3f}$ for $T^R _{3f}=0$ (actually, this is true for any field that transforms as a singlet by the SM symmetry regardless of its chirality), then,

\begin{widetext}
\begin{equation}
\begin{split}
\bar{\Psi} ^R i \gamma ^\mu D_\mu ^R \Psi ^R& = - e Q_f \bar{\psi} _f ^R \gamma ^\mu \psi _f ^R A_\mu\\
&- \left[ - g_Z Q_f \sin ^2 \theta _W \cos \xi - \frac{1}{2} \left( \epsilon g _Z Q_{Yf} ^R \tan \theta _W + g_X Q_{Xf} ^R \right) \sin \xi \right] \bar{\psi} _f ^R \gamma ^\mu \psi _f ^R Z_\mu \\
&- \left[ - g_Z Q_f \sin ^2 \theta _W \sin \xi + \frac{1}{2} \left( \epsilon g _Z Q_{Yf} ^R \tan \theta _W + g_X Q_{Xf} ^R \right)\cos \xi \right] \bar{\psi} _f ^R \gamma ^\mu \psi _f ^R Z' _\mu.
\end{split}
\label{diracRu1}
\end{equation}
\end{widetext}

The sum of Eq.\eqref{diracLu1} and Eq.\eqref{diracRu1} accounts for the overall interactions between left and right-handed fermions with the physical neutral gauge bosons. It is useful to separate the contributions for when $Q_{Xf} ^{L,R}=0$ and for $Q_{Xf} ^{L,R} \neq 0$ 

\begin{widetext}
\begin{equation}
\begin{split}
\mathcal{L} = &- e Q_f \bar{\psi} _f \gamma ^\mu \psi _f A_\mu \\
&- \left[ g_Z \left( T_{3f} - Q_f \sin ^2 \theta _W \right) \cos \xi - \frac{1}{2} \epsilon g _Z Q_{Yf} ^L \tan \theta _W \sin \xi \right] \bar{\psi} _f ^L \gamma ^\mu \psi _f ^L Z_\mu \\
&- \left[ - g_Z Q_f \sin ^2 \theta _W \cos \xi - \frac{1}{2} \epsilon g _Z Q_{Yf} ^R \tan \theta _W \sin \xi \right] \bar{\psi} _f ^R \gamma ^\mu \psi _f ^R Z_\mu \\
&- \left[ g_Z \left( T_{3f} - Q_f \sin ^2 \theta _W \right) \sin \xi + \frac{1}{2} \epsilon g _Z Q_{Yf} ^L \tan \theta _W \cos \xi \right] \bar{\psi} _f ^L \gamma ^\mu \psi _f ^L Z' _\mu \\
&- \left[ - g_Z Q_f \sin ^2 \theta _W \sin \xi + \frac{1}{2} \epsilon g _Z Q_{Yf} ^R \tan \theta _W \cos \xi \right] \bar{\psi} _f ^R \gamma ^\mu \psi _f ^R Z' _\mu \\
&+ \frac{1}{2} g_X Q_{Xf} ^L \sin \xi \bar{\psi} _f ^L \gamma ^\mu \psi _f ^L Z_\mu + \frac{1}{2} g_X Q_{Xf} ^R \sin \xi \bar{\psi} _f ^R \gamma ^\mu \psi _f ^R Z_\mu - \frac{1}{2} g_X Q_{Xf} ^L \cos \xi \bar{\psi} _f ^L \gamma ^\mu \psi _f ^L Z' _\mu \\
&- \frac{1}{2} g_X Q_{Xf} ^R \cos \xi \bar{\psi} _f ^R \gamma ^\mu \psi _f ^R Z' _\mu. \\ \\
\end{split}
\label{generaldiracu1}
\end{equation}
\end{widetext}

The last two lines of Eq.\eqref{generaldiracu1} are the contributions introduced when the charges of the fermions under $U(1)_{X}$ are non-zero. In the limit we are working, $M_{Z^{\prime}} \ll M_{Z}$, i.e. with $\cos \xi \sim 1$, the interactions mediated by the standard $Z$ boson are identical to the SM case, even in the case when $Q_{Xf} ^{L,R} \neq 0$. Thus we get,

\begin{equation}
\mathcal{L} _{Z} = - \left( g_Z J^\mu _{NC} \right) Z _\mu.
\end{equation}

For the $Z^{\prime}$ boson we have two contributions, the first one for when $Q_{Xf} ^{L,R} = 0$,

\begin{equation}
\mathcal{L} _{Z'} = - \left( \epsilon e J^\mu _{em} + \epsilon _Z g_Z J^\mu _{NC} \right) Z' _\mu,
\label{Zprime1}
\end{equation}

and a second one, which exists when $Q_{Xf} ^{L,R} \neq 0$,

\begin{equation}
\mathcal{L} _{Z'} = - \left( \frac{1}{2} g_X Q_{Xf} ^L \bar{\psi} _f ^L \gamma ^\mu \psi _f ^L
+ \frac{1}{2} g_X Q_{Xf} ^R \bar{\psi} _f ^R \gamma ^\mu \psi _f ^R \right) Z' _\mu.
\label{Zprime2}
\end{equation}

Notice that Eq.\eqref{Zprime1} represent the well-known DarK Z boson interactions, and its implications on parity violation, rare decays and Higgs physics have been studied in \cite{Davoudiasl:2012ag}. Moreover, when the mass mixing between the hidden boson and the Z boson is neglected in Eq.\eqref{Zprime1}($\epsilon_{z}=0$), we fall back to the Dark photon model, where the $Z^{\prime}$ couples to the SM particles proportionally to the kinectic mixing $\epsilon e$. Consequently, we are studying a more general version of the models already studied in the literature. The next and last step that we will take in this section is to write the interaction among the $Z^{\prime}$ boson and SM fermions in the form,

\begin{eqnarray}\label{LintZp}
\,{\cal L}_{Z^{\prime}}^{\, int}= g_{V}^{(\Psi_{i})} \, \, \overline{\Psi_{i}} \, \slashed{Z}^{\prime} \, \Psi_{i}
+ g_{A}^{(\Psi_{i})} \, \, \overline{\Psi_{i}} \, \slashed{Z}^{\prime} \, \gamma_{5} \, \Psi_{i} \; ,
\end{eqnarray}with $\Psi_{i}$ being each one of the fermions of the SM, $g^{V}$ and $g^{A}$ the vector and axial couplings which are set by the $U(1)_X$ charges, which characterizes each one of the anomaly free 2HDM-$U(1)_X$ models of the table \ref{Table1}. Focusing on the charged leptons, Eq.\eqref{LintZp} can be written as,
\begin{widetext}
\begin{equation}
\mathcal{L}_{Z'} = \overline{e}_{i} \, \slashed{Z} \left\{\epsilon \, e + \frac{\epsilon_{Z}g_{Z}}{4}(1-4\sin ^2 \theta _W)+\frac{g_{X}}{8}(7u+5d)+\left[-\frac{\epsilon_{Z}g_{Z}}{4}+\frac{g_{X}}{8}(u-d)\right]\gamma_5\right\}e_{i} \; .
\label{currentgeralZprimeleptons}
\end{equation}
\end{widetext}
The corresponding lagrangian for the light neutrinos $\nu_{i}\,(i=1,2,3)$ is,

\begin{equation}
\mathcal{L}_{Z'} = \overline{\nu}_{i} \, \slashed{Z}\left\{\frac{3g_{X}}{8}(u+d)-\frac{g_{Z}\epsilon_{Z}}{4}+\left[-\frac{3g_{X}}{8}(u+d)+\frac{g_{Z}\epsilon_{Z}}{4}\right]\gamma _5\right\}\nu_i \; .
\label{currentgeralZprimeneutrinos}
\end{equation}
Lastly, for the quarks with positive and negative isopin, we find,
\begin{widetext}
\begin{equation}
\mathcal{L}_{Z'}  = \overline{u}_i \, \slashed{Z} \left\{-\frac{2}{3}\epsilon e - \frac{\epsilon_{Z}g_{Z}}{4}\left(1-\frac{8}{3}\sin^2 \theta_W\right)-\frac{g_{X}}{8}(3u+d)+\left[\frac{\epsilon_{Z}g_{Z}}{4}+\frac{g_{X}}{8}(d-u)\right]\gamma _5\right\}u_i \; .
\label{currentgeralZprimeupquarks}
\end{equation}
\end{widetext}
\begin{widetext}
\begin{equation}
\mathcal{L}_{Z'} = \overline{d}_i\, \slashed{Z} \left\{\frac{1}{3}\epsilon \, e + \frac{\epsilon_{Z}g_{Z}}{4}\left(1-\frac{4}{3}\sin ^2 \theta_W\right)-\frac{g_{X}}{8}(u+3d)+\left[-\frac{\epsilon_{Z}g_{Z}}{4}+\frac{g_{X}}{8}(u-d)\right]\gamma _5\right\}d_i \; .
\label{currentgeralZprimedownquarks}
\end{equation}
\end{widetext}
With Eq.\eqref{currentgeralZprimeleptons}, Eq.\eqref{currentgeralZprimeneutrinos}, Eq.\eqref{currentgeralZprimeupquarks}, and Eq.\eqref{currentgeralZprimedownquarks}, we can derive the $Z^\prime$ branching ratio into any fermions.  Using these equations, we computed the total decay width as well as the branching ratio into neutrinos and $e^+e^-$.

\bibliography{references}

\end{document}